\documentclass{elsart}
\usepackage{epsfig}
\begin{document}
\begin{flushright}
ADP-01-46/T478\\
\end{flushright}
\begin{frontmatter}
\title{Low-lying eigenmodes of the Wilson-Dirac operator
	and correlations with topological objects}
\author{Daniel-Jens Kusterer\thanksref{DJK}},
\author{John Hedditch\thanksref{JH}},
\author{Waseem Kamleh\thanksref{WK}}, 
\author{Derek B. Leinweber\thanksref{DBL}} and 
\author{Anthony G. Williams\thanksref{AGW}}
\thanks[DJK]{E-Mail: djkuster@physics.adelaide.edu.au}
\thanks[JH]{E-Mail: jhedditc@physics.adelaide.edu.au}
\thanks[WK]{E-Mail: wkamleh@physics.adelaide.edu.au}
\thanks[DBL]{E-Mail: dleinweb@physics.adelaide.edu.au}
\thanks[AGW]{E-Mail: awilliam@physics.adelaide.edu.au}
\address{CSSM Lattice Collaboration, Special Research Center for the Subatomic Structure of Matter
(CSSM) and Department of Physics and Mathematical Physics, University
of Adelaide 5005, Australia}
\begin{abstract}
The probability density of low-lying eigenvectors of the hermitian Wilson-Dirac operator $H(\kappa )=\gamma_5 D_{\mathrm{W}}(\kappa )$ is examined. Comparisons in position and size between eigenvectors, topological charge and action density are made. We do this for standard Monte-Carlo generated SU(3) background fields and for single instanton background fields. Both hot and cooled SU(3) background fields are considered. An instanton model is fitted to eigenmodes and topological charge density and the sizes and positions of these are compared.
\end{abstract}
\end{frontmatter}

\section{Introduction}

It is known \cite{SST98,SS97,EHN98,EHN982,JLSS96} that the zero modes
of lattice Dirac operators are strongly localised. We show that not
only zero modes but all low-lying modes of the hermitian Wilson-Dirac
operator $H(\mathrm{m}_0)=\gamma_5 D_{\mathrm{W}}(-\mathrm{m}_0)$ are
strongly localised for bare quark masses $0\leq\mathrm{m}_0\leq
2$. Furthermore, we show that this localization is strongly correlated
to topological objects, including instantons. A similar examination
has been done for the overlap formalism \cite{DH00}. We are interested
in examining the extent to which similar properties are already
manifested at the kernel level of the overlap formalism.

Our Wilson-Dirac operator is defined in the standard way by
\begin{eqnarray}
[ D_{\mathrm{W}}(\kappa)\psi ] (x) & = & \psi(x)-\kappa \sum_\mu [ 
(1-\gamma_{\mu})U_{\mu}(x)\psi(x+\mu)\nonumber\\
        & & +(1+\gamma_{\mu})U_{\mu}^{\dagger}(x-\mu)\psi(x-\mu) ],
\label{eq:Dwilson}
\end{eqnarray}
where the hopping parameter $\kappa$ is related to the bare mass by
\begin{equation}
\kappa = (8-2m_0)^{-1}.
\end{equation}
Therefore the above mentioned bare mass range is equivalent to
$0.125\leq\kappa\leq 0.25$ at tree level. We examine eigenmodes in and
just outside the region $m_c < m_0 < 2$, where $m_c$ is the
``critical-mass'' which is 0 at tree-level but for non-trivial gauge
fields shifts away from 0. This is the range of the mass parameter
$m_0$ needed for use in the overlap formalism. Modes of $H(m_0)$
crossing zero in this region are accompanied by the abrupt appearance
of exact zero modes of the overlap-Dirac operator at the corresponding
$m_0$ value \cite{EHN99}.

We solve the eigenvalue problem $D_{\mathrm{W}}\psi(x)=\lambda\psi(x)$
for the first four low-lying eigenvalues. We are interested in
calculating eigenmodes in the physical region of the overlap formalism
with some further points just at the edge of this region. This
corresponds to a $\kappa$-region starting slightly below the critical
value of $\kappa = \kappa_{\mathrm{c}}$ and extending slightly beyond
the point where doublers appear in the continuum limit of the overlap
formalism. We consider $0.115\leq\kappa\leq0.26$ at tree level. Since
the critical $\kappa$-value shifts from its free field value of
$0.125$ to some higher value for non-zero gauge coupling, we have to
adjust our $\kappa$-range accordingly. We look for any change in
behaviour of the eigenmodes at the border of the region of physical
interest. 

Eigenmodes are found by an accelerated conjugate gradient routine
\cite{CG} which is further improved by using dynamic state
renormalisation. The major advantage of a conjugate gradient algorithm
besides its almost perfect parallel structure is that it yields not
only eigenvalues with appropriate degeneracies, but eigenvectors, as
well. For selected $\kappa$-values we also calculate up to 20
eigenmodes. In the following, the phrase ''low-lying eigenmodes''
should be understood to mean eigenmodes corresponding to the low-lying
eigenvalues.

In order to examine localisations of calculated eigenmodes we plot the
probability density $\rho(x)=\parallel\psi(x)\parallel^2$ for three
dimensional cuts through the lattice. For comparison we plot the
action density as well as the topological charge density for the
appropriate configuration in the same way. 
Our calculations are made on $8^3\times 16$, $16^4$ and $16^3\times 32$ lattices with anti-periodic fermion boundary conditions in the time direction. On the $8^3\times 16$ lattice we test the correlation of low-lying eigenmodes with a single instanton configuration and standard Monte-Carlo generated SU(3) background fields. For the latter background fields we consider hot, 5-sweep and 12-sweep cooled configurations. Calculations on the larger lattices are to verify the conclusions we obtain from the smaller lattice. In order to quantify our results further we fit an instanton model to the obtained data. We distinguish between the model for action or charge densities and the model for the zero-mode density \cite{GGLRS},
\begin{equation}
p(x)_{\mathrm{act}} = c \cdot\frac{6}{\pi^2}\cdot \frac{\rho^4}{\left( (x-x_0)^2+\rho^2\right)^4},\label{eq:instActModel}
\end{equation}
\begin{equation}
p(x)_{\mathrm{zero}} = c \cdot\frac{2}{\pi^2}\cdot \frac{\rho^2}{\left( (x-x_0)^2+\rho^2\right)^3},\label{eq:instZeroModel}
\end{equation}
where $x$ is the distance from the instanton peak at $x_0$ to the
calculated densities. The normalisation factor $c$ allows us to fit to
the instanton shape and prevents the fit from being dominated by the
maximal value of the fitted object which is affected by periodic
images due to the finite volume of the 4-torus. Both models are
continuum results derived from the standard 't Hooft
ansatz. Eq.~(\ref{eq:instActModel}) is the action density and is used
to fit action and charge densities. Eq.~(\ref{eq:instZeroModel}) is
the density of the fermion field in the zero-mode and is used to fit
eigenmode densities. This allows us to compare localisation sizes and
positions quantitatively.

This paper is organised as follows: After a discussion in
Sec.~\ref{sect:lattice} of lattice techniques and general properties,
we describe results for a single instanton configuration on a $16^4$
lattice in Sec.~\ref{sect:inst}. In Sec.~\ref{sect:su3} we investigate
localisations of eigenmodes of the hermitian Wilson-Dirac operator and
the correlations those eigenmodes have with topology on standard
Monte-Carlo generated SU(3) configurations on a $16^3\times 32$
lattice. We do this for both hot and 12-sweep-cooled
configurations. We quantify those localisations and their correlations
to topology and compare sizes of the eigenmode localization with sizes
of the corresponding topological object in
Sec.~\ref{sect:quant}. Conclusions are presented in
Sec.~\ref{sect:concl}.

\section{Lattice techniques}\label{sect:lattice}
We carry out our studies on three different lattices. Single instanton
configurations are created on an $8^3\times 16$ lattice with $\rho =
1.0$ and on a $16^4$ lattice with $\rho = 2.0$. The latter
configuration is cooled for 2 sweeps using the standard Wilson action
to minimize boundary effects. We consider three standard Monte-Carlo
SU(3) configurations generated on an $8^3\times 16$ lattice with
$\beta = 4.38$, which corresponds to a lattice spacing of 0.165(2)
fm. The second lattice size used is $16^3 \times 32$ with $\beta =
4.60$ which corresponds to a lattice spacing of 0.125(2) fm. These
configurations are generated with a plaquette plus rectangle improved
action with mean-field improved coefficients. For the smaller
exploratory lattice the standard Wilson action and standard plaquette
topological charge are used to minimize boundary effects. For the
bigger lattice, a three-loop improved action and three-loop improved
topological charge operator \cite{SBT,deF} are used.

The topological charge density is given by
\begin{equation}
q(x) = \frac{g^2}{32\pi ^2} \epsilon_{\mu\nu\rho\sigma}
{\mathrm{Tr}}(F_{\mu\nu}(x)F_{\rho\sigma}(x)).
\end{equation}
For the standard plaquette topological charge we use the traceless
definition of $F_{\mu\nu}$ extracted from the consideration of
$1 \times 1$ plaquettes alone \cite{FB}
\begin{equation}
F_{\mu\nu} = {-i\over{8}} \left[ W^{1 \times 1} - 
             W^{1 \times 1^{\dagger}} \right]_{\rm Traceless}
\end{equation}
where $W^{1 \times 1}$ is the clover-sum of four $1 \times 1$ Wilson
loops lying in the $\mu ,\nu$ plane.  For the three-loop improved
topological charge operator, we employ an ${\mathcal O}(a^4)$ improved
definition of $F_{\mu\nu}$ in which the standard clover-sum of four $1
\times 1$ Wilson loops lying in the $\mu ,\nu$ plane is combined with
$2 \times 2$ and $3 \times 3$ Wilson loop clovers.  Bilson-Thompson
{\it et al.} \cite{SBT} find
\begin{equation}
F_{\mu\nu} = {-i\over{8}} \left[\left( {3\over{2}}W^{1 \times 1}-{3\over{20u_0^4}}W^{2 \times 2}
+{1\over{90u_0^8}}W^{3 \times 3}\right) - {\mathrm{h.c.}}\right]_{\rm
Traceless}
\end{equation}
where $W^{n \times n}$ is the clover-sum of four $n \times n$ Wilson
loops and where $F_{\mu\nu}$ is made traceless by subtracting $1/3$ of
the trace from each diagonal element of the $3 \times 3$ colour
matrix.  This definition reproduces the continuum limit with ${\mathcal
O}(a^6)$ errors.  On approximately self-dual configurations, this
operator produces integer topological charge to better than 4 parts
in $10^4$.

We use periodic boundary conditions in space and anti-periodic in the
time direction for fermions. Each of the hot configurations is also
cooled for 12 sweeps using a Cabibbo Marinari based algorithm in which
the three diagonal SU(2) subgroups of SU(3) are looped over twice
\cite{FB} using a three-loop ${\mathcal O}(a^4)$ improved action
\cite{SBT,deF}. Twelve sweeps of cooling is just enough to see clear
structure on the topological charge and action densities, but it still
preserves much of the original topology. To see correlations for some
eigenmode densities we also need to use the topological charge density
of 5-sweep cooled configurations. These topological charge densities
are much rougher, but are closer to the original configuration, with
topological objects moving less than in the 12-sweep cooled
configurations.

The visualisations of the eigenmode probability densities are treated
consistently in order to allow direct comparison. This means they are
all normalised with respect to the maximum value of each eigenmode. In
this way, an isosurface at half the peak height will reflect the size
of the object.

\section{Smooth instanton background}\label{sect:inst}

We generate a single instanton background on an $8^3 \times 16$ and on
a $16^4$ lattice by performing the path integration of
\begin{equation}
A_{\mu}(x) = \frac{x^2}{x^2+\rho^2}\left(\frac{i}{g}\right) (\partial_{\mu}S) S^{-1},
\end{equation}
with
\begin{equation}
S = \frac{x_4 \pm i \overrightarrow{x} \cdot \overrightarrow{\sigma}}{\sqrt{x^2}},
\end{equation}
where $+$ is an instanton and $-$ an anti-instanton, to create the
link variable. We find in the regular gauge
\begin{equation}
U^{\rm reg}_\mu(x) = \exp \Bigl[ia_\mu(x)\cdot\sigma \phi_\mu(x;\rho)\Bigr],
\end{equation}
\begin{equation}
\phi_\mu(x;\rho) = {1\over \sqrt{ \rho^2 + \sum_{\nu\ne\mu} (x_\nu-\overline{x}_\nu)^2}}
\tan^{-1} { \sqrt{ \rho^2 + \sum_{\nu\ne\mu} (x_\nu-\overline{x}_\nu)^2} \over
\rho^2 + \sum_{\nu} (x_\nu-\overline{x}_\nu)^2 + (x_\nu-\overline{x}_\nu) },
\end{equation}
\begin{eqnarray}
a_1(x) & = & (-x_4+\overline{x}_4, x_3-\overline{x}_3, -x_2+\overline{x}_2),\cr
a_2(x) & = & (-x_3+\overline{x}_3, -x_4+\overline{x}_4, x_1-\overline{x}_1),\cr
a_3(x) & = & (x_2-\overline{x}_2, -x_1+\overline{x}_1, -x_4+\overline{x}_4),\cr
a_4(x) & = & (x_1-\overline{x}_1, x_2-\overline{x}_2, x_3-\overline{x}_3).
\end{eqnarray}
In the singular gauge we find\footnote{Note that the analogous result
of \cite{EHN98} inverts the roles of instantons and anti-instantons}
\begin{equation}
U^{\rm sing}_\mu(x) = \exp \Bigl[ib_\mu(x)\cdot\sigma 
\bigr(\phi_\mu(x;0)-\phi_\mu(x;\rho)\bigl)\Bigr],
\end{equation}
\begin{eqnarray}
b_1(x) & = & (x_4-\overline{x}_4, x_3-\overline{x}_3, -x_2+\overline{x}_2),\cr
b_2(x) & = & (-x_3+\overline{x}_3, x_4-\overline{x}_4, x_1-\overline{x}_1),\cr
b_3(x) & = & (x_2-\overline{x}_2, -x_1+\overline{x}_1, x_4-\overline{x}_4),\cr
b_4(x) & = & (-x_1+\overline{x}_1, -x_2+\overline{x}_2, -x_3+\overline{x}_3).
\end{eqnarray}

\begin{figure}[p]
\epsfig{figure=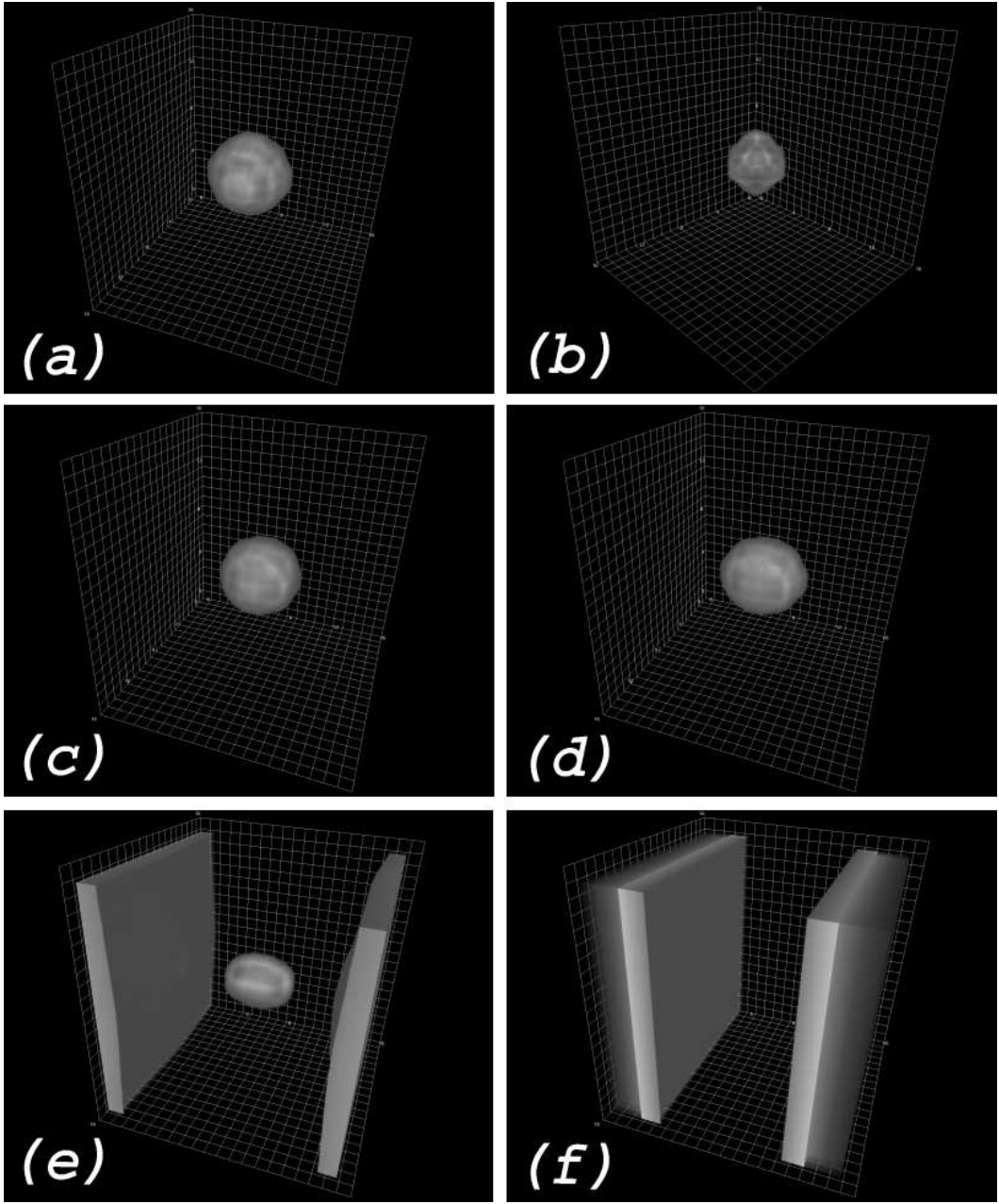}
\caption{(a) Action density of a single instanton configuration on a $16^4$ lattice. (b) First eigenmode for the single instanton configuration for $\kappa = 0.19$. Note the very strong correlation with the instanton on the action density.
(c)  First eigenmode for the single instanton configuration for $\kappa = 0.25$. Again note the strong correlation between the eigenmode and the action density.
(d) Second eigenmode for the single instanton configuration for $\kappa = 0.25$. The localization has a prolate shape compared to the spherical instanton.
 (e) Second eigenmode for the single instanton configuration for $\kappa = 0.19$. The localization has a wall like shape with a prolate correlation to the instanton. 
(f) Third eigenmode for the single instanton configuration for $\kappa = 0.19$. The eigenmode has a wall like shape and is not correlated with the instanton. For all figures what is shown is a volume rendering of the corresponding density. The outer isosurface is half the peak density.}
\label{fig:instAll}
\end{figure}

The singular gauge instanton is clearly recognisable on the volume
rendered action density plot as seen in Fig.\ \ref{fig:instAll}
(a). The outer surface shown is half the peak height.

\begin{figure}[tbp]
\epsfig{angle=90, file=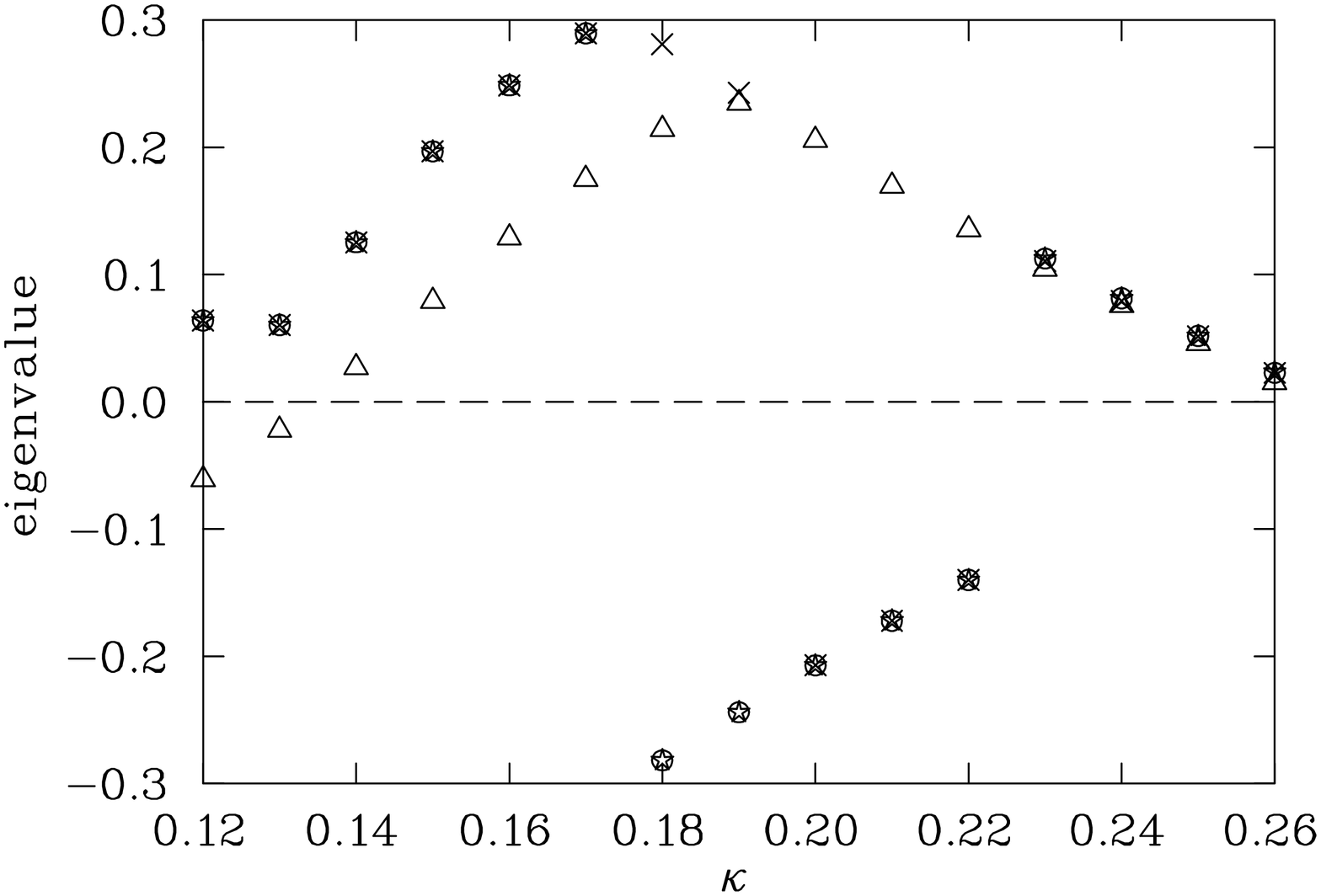, height=9.0cm}
\caption{Smallest four eigenvalues of the eigenvalue spectrum of a smooth single instanton configuration on a 2-sweep cooled $16^4$ lattice. For the open triangles we have strong spherical correlation between the eigenmode and action density and behaviour as described in the text. For the other symbols we have correlation with prolate-like shapes and wall-like structures. The shapes are shown in Fig.\ \ref{fig:instAll}. For $\kappa = 0.20, 0.21$ and $0.22$ we expect the fifth eigenvalue to lie degenerate with the open triangles.}
\label{fig:instSpectr}
\end{figure}

The results seen on the smaller exploratory lattice are also found on
the $16^4$ lattice. The instanton on the latter lattice is cooled for
two sweeps to minimise boundary effects.  Eigenmodes of this
configuration are calculated for $0.12\le\kappa\le 0.27$ with an
increment of 0.01 between values. The first four eigenvalues of the
spectrum are shown in Fig.\ \ref{fig:instSpectr}. We then evaluate the
localisation of $\rho(x)$ for the first three low-lying eigenmodes for
each $\kappa$. This is done by plotting $\rho(x)$ as seen in Fig.\
\ref{fig:instAll} for selected eigenmodes.

For eigenmodes with $\kappa\ll
\kappa_{\mathrm{c}}=0.125$ we find a rather uniform distribution of $\rho(x)$ whereas eigenmodes for all $\kappa_{\mathrm{c}}<\kappa<0.27$ are strongly localised. For the lowest eigenmode this localisation strongly corresponds to the localisation of the instanton in the action density plot as seen by comparing Fig.\ \ref{fig:instAll} (a) and Fig.\ \ref{fig:instAll} (b), (c). We therefore say the eigenmode displays strong correlation with the instanton. The size and shape of this correlation varies from broad with some wall like structures for $\kappa\le\kappa_{\mathrm{c}}$ to very small for $\kappa_{\mathrm{c}}<\kappa<0.19$ and is getting broader again for larger values of $\kappa$. For higher eigenmodes this localization gets broader and less correlated with the instanton. Some higher eigenmodes show no correlation with the instanton but just wall like structure as seen in Fig.\ \ref{fig:instAll} (f). Others show prolate-like correlations as seen in Fig.\ \ref{fig:instAll} (d). Some eigenmodes also show a wall-like structure and a prolate-like correlation as seen in Fig.\ \ref{fig:instAll} (e). It is useful to note that plane wave behaviour would display uniform behaviour in the density plot.

\section{Monte-Carlo generated SU(3) gauge field background}\label{sect:su3}
\begin{figure}[p]
\epsfig{figure=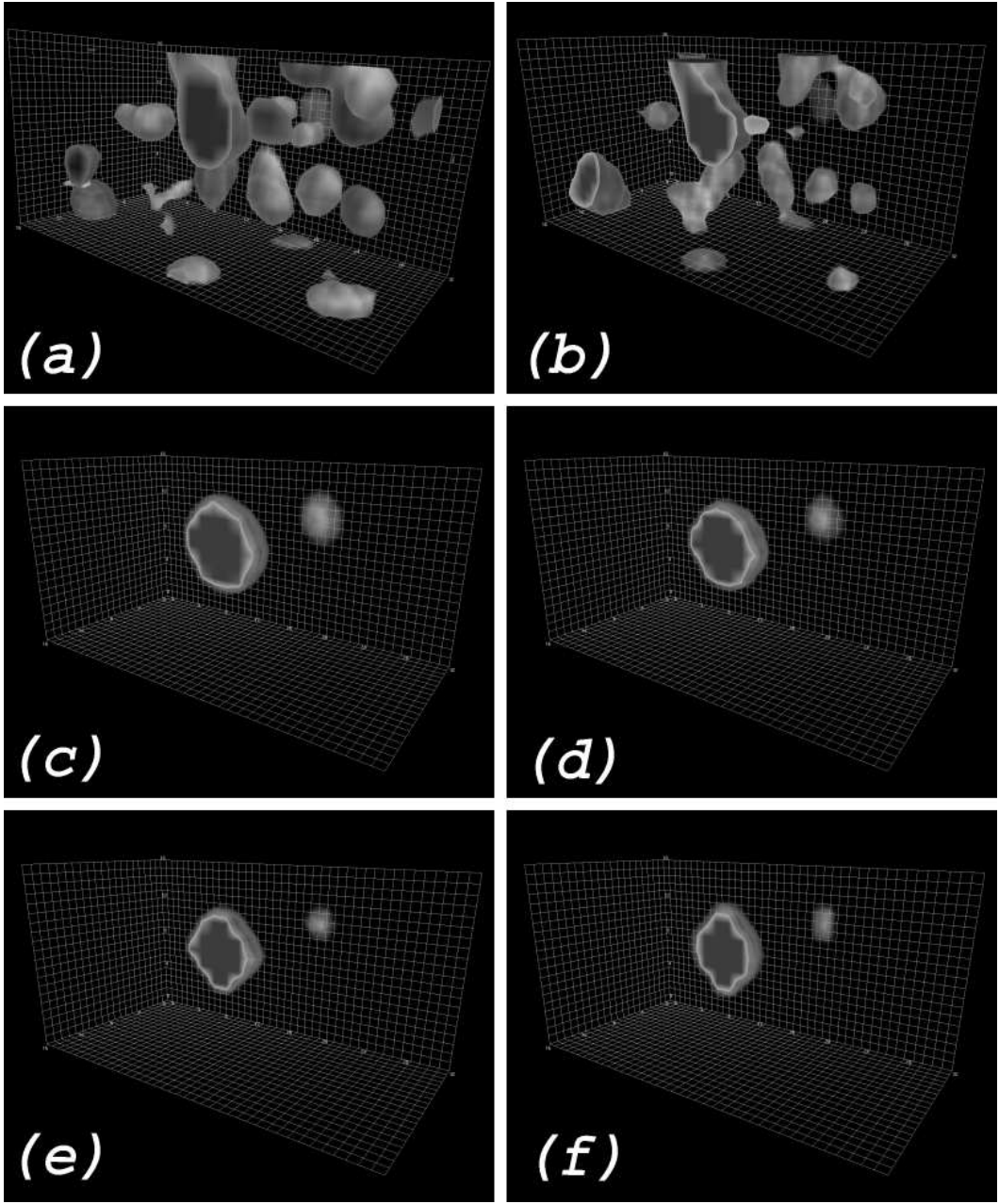}
\caption{(a) Topological charge density of a 12-sweep cooled
$16^3\times 32$ configuration at $\beta = 4.60$. (b) Action density of
the same configuration. There is clearly less structure for the action
density than the topological charge density. (c) The low-lying
eigenmode density at $\kappa = 0.14$ of the same configuration with
correlation to the object seen in the action density and the
topological charge density. (d) Eigenmode density at $\kappa = 0.16$
showing correlation to the same topological object. (e) Eigenmode
density at $\kappa = 0.18$. The localization is maximal for this value
of $\kappa$. (f) Eigenmode density at $\kappa = 0.19$. The
localization is becoming larger again.}
\label{fig:su3All}
\end{figure}

After seeing that the low-lying eigenmodes of the Wilson-Dirac
operator are strongly correlated to the instanton on a one-instanton
configuration, we next carry out investigations on standard
Monte-Carlo generated SU(3) background fields. We investigate three
hot and three cooled configurations on both $8^3\times 16$ ($\beta =
4.38$) and $16^3 \times 32$ ($\beta = 4.60$) lattices. Three-loop
improved cooling \cite{SBT,deF} is used to cool the configurations for
12 sweeps. This is just enough to get clear structure in the action
and topological charge. As a reference for comparison of the eigenmode
density we use the action and topological charge density of the cooled
configuration. We do this because those densities are too rough for
the hot configuration and no accurate comparison would be possible. As
there is more structure observed on the topological charge density
plot, we use this as our preferred reference. See Fig.\
\ref{fig:su3All} (a) for a typical topological charge density plot of
a $16^3 \times 32$ configuration. However the action density plot is
also useful for guidance. 

We say an eigenmode is correlated to a topological object if $\rho(x)$
has a peak within one lattice site of the peak topological charge
density. We calculate the first four eigenmodes for values of
$0.13\le\kappa\le 0.25$ for cooled configurations and
$0.15\le\kappa\le 0.29$ for hot configurations in steps of 0.01. This
range is from approximately 0.02 smaller than $\kappa_{\mathrm{c}}$ to
a region where doublers appear in the overlap formalism. We find that
the behaviour described in the following is general for all
configurations: Each of the eigenmodes is localised. This was already
observed for the lowest eigenmode \cite{JLSS96}. For
$\kappa<\kappa_{\mathrm{c}}$ this localization weakens and the
density, $\rho (x)$, broadens quickly. An exponential decay of the
density $\rho(x)$, as previously observed \cite{SS97,JLSS96}, seems
likely to occur. All low-lying eigenmodes for
$\kappa\ge\kappa_{\mathrm{c}}$ are correlated to topological
objects.

\begin{figure}[tbp]
\epsfig{angle=90, figure=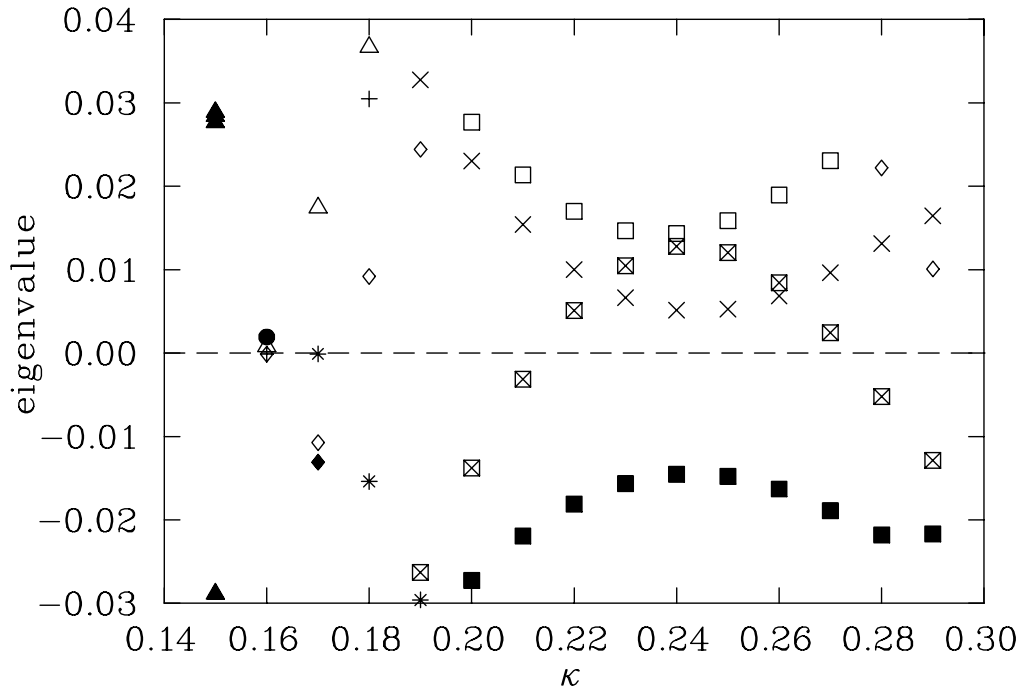, height=9.0cm}
\caption{Smallest four eigenvalues of the eigenvalue spectrum of a hot configuration showing correlations of eigenmodes to topological objects. Each eigenmode has a correlation to one topological object and each different symbol indicates a different topological object. Modes for $\kappa = 0.15 < \kappa_{\mathrm{c}}$ are very broad and weakly correlated to several objects.}
\label{fig:modes}
\end{figure}

We are able to track a correlation along the modes for all
$\kappa\ge\kappa_{\mathrm{c}}$ and for one step smaller than
$\kappa_{\mathrm{c}}$. Thus we can label a mode by its
correlation. Fig.\ \ref{fig:modes} shows eigenmodes of a hot
configuration, where the symbol used denotes which topological object
the localised low-lying eigenmode is correlated with. Careful
inspection of Fig.~\ref{fig:modes} reveals the presence of an
eigenmode, correlated to one topological object, but with a spectral
flow containing two zero crossings. The size of the correlated object,
obtained in the following section, suggests that this object is a
lattice artifact.

Around $\kappa_{\mathrm{c}}$ and towards the upper end of the analysed
spectrum, localisations are weak. However, the correlation of
positions is strong. At these $\kappa$ we tend to get correlations to
more than one topological object. The localisations get stronger and
sharper for increasing values of $\kappa$ until a maximal localization
is reached for a value $\kappa$ we will define to be
$\kappa_{\mathrm{max}}$. For $\kappa>\kappa_{\mathrm{max}}$ the
localisations get weaker again. Fig.\ \ref{fig:su3All} shows a
visualisation of this behaviour. Additional figures can be found
elsewhere \cite{web}. We take a closer look at this behaviour in the
next section.

For the lower eigenmodes, which are separated by a gap from the higher
eigenmodes, we find correlation to one topological object per
eigenmode. For higher eigenmodes , which are closer together with some
degenerate modes, we find correlations to more than one topological
object. In general we find that the closer the eigenvalues and the
weaker the actual localisations, the more likely it is to get
correlation with more than one topological object in the corresponding
eigenmodes.

Calculations of 20 eigenmodes for selected values of $\kappa$ on hot
configurations show only little broadening in the localisations and
persistent correlations with topology. This suggests that this
behaviour will not change quickly and such correlations will persist
for even higher eigenmodes. 

As mentioned above, we use the topological charge density of the
12-sweep cooled configuration for comparison. The position of
eigenmodes on cooled fields agree perfectly well with the position of
topological objects seen in this density. A typical distance between
such positions being 0.05 lattice units. For hot configurations some
of the correlations are slightly offset compared to the structure in
the smoothed configurations whereas other structure in the eigenmodes
has no corresponding structure in the smoothed topological charge. But
in those cases a comparison with a less cooled configuration reveals
correlations between eigenmodes and topological objects which are
moved or destroyed by further cooling. This behaviour is expected and
understood, as topological objects are known to move under cooling as
instantons and anti-instantons attract each other and annihilate when
they are close enough together.

This clear correlation between $\rho(x)$ of eigenmodes on hot
configurations and topological objects suggests that it is possible to
identify areas through the noise of a hot configuration with
significant topology. We can ``see through'' the noise by using
eigenmodes of the hermitian Wilson-Dirac operator.

\begin{figure}[tbp]
\epsfig{angle=90, figure=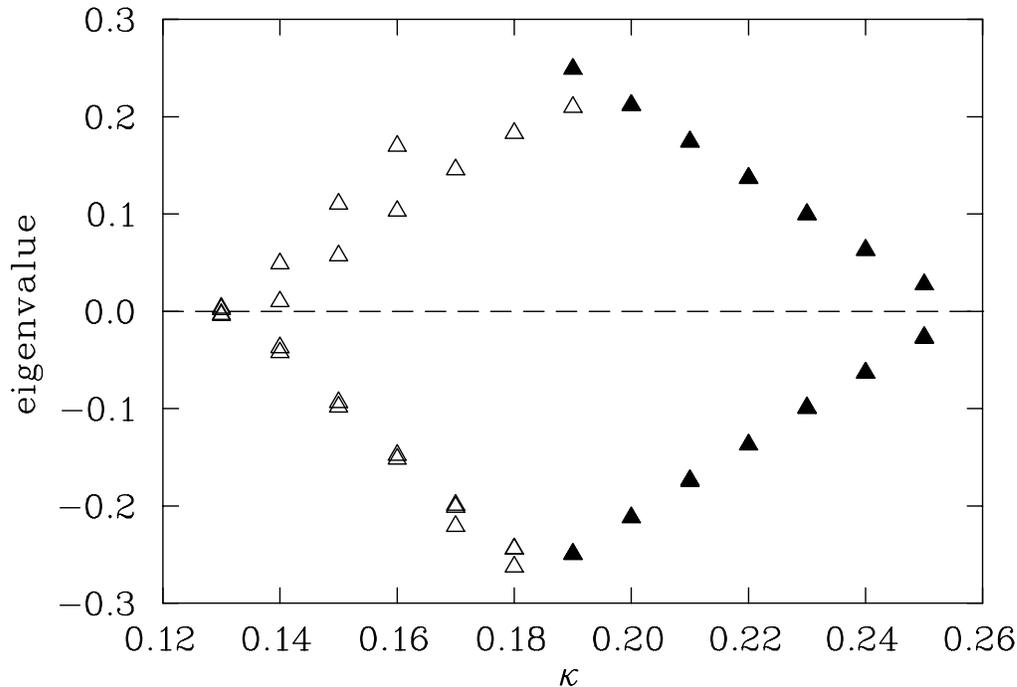, height=9.0cm}
\caption{First four eigenvalues of the spectrum of a 12-sweep cooled
configuration. Note the rhomboid shape with the area without
eigenvalues. The modes to the right of the maximum, the filled
symbols, are only weakly localised. The open symbols are strongly
localised.}
\label{fig:coolSpectr}
\end{figure}

The spectral plot of the first four eigenvalues of a 12-sweep cooled
configuration looks different compared to the same plot on a hot
configuration. Comparing Fig.~\ref{fig:modes} and
Fig.~\ref{fig:coolSpectr} we notice the rhomboid shape with an area
without any eigenvalues in the spectral plot of the cooled
configuration. This is expected for such smoothed configurations
\cite{NL}. Eigenmodes on the right-hand side of this rhomboid behave
different than eigenmodes on the left-hand side of the rhomboid. The
localization of the eigenmodes on the left-hand side is approximately
the same as the localisations in the hot configuration. But the
eigenmodes on the right-hand side are much weaker in
localization. Those weaker localised eigenmodes are related to very
high eigenmodes in the hot configuration and the process of cooling
brings them into an area where we can observe them as low
eigenmodes. These eigenmodes must be very high eigenmodes in a hot
configuration for they are much weaker localised as localisations we
have seen for up to the 20th eigenmode in a hot
configuration. Although the behaviour in localization strength for
eigenmodes of cooled configurations is different from that of
eigenmodes of hot configurations the correlations with topological
objects still exist for all eigenmodes. We will take a closer look at
localization strength and quantify it in the next section.

\section{Quantitative Results}\label{sect:quant}

\begin{figure}[tbp]
\epsfig{angle=90, figure=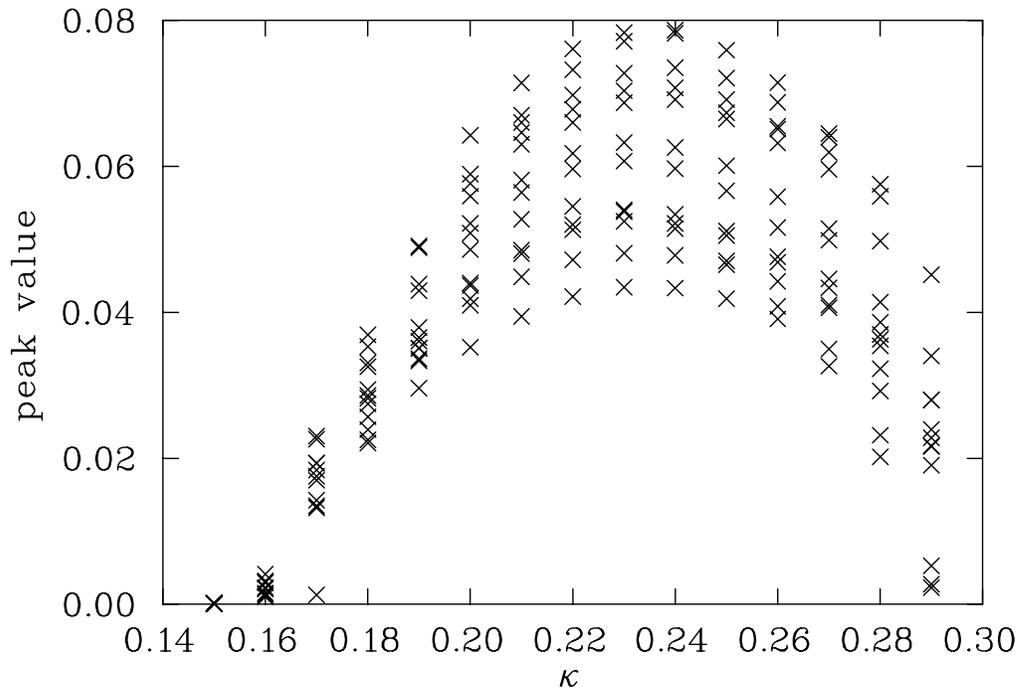, height=9.0cm}
\caption{Peak value of four eigenmode densities each from hot
$16^3\times32$ configurations with respect to $\kappa$}
\label{fig:maxHot}
\end{figure}

As described earlier in this paper, the localization of the eigenmodes
change shape and size with changing of $\kappa$. In order to quantify
this behaviour we have two methods. The first one is fairly simple. We
have already seen that most eigenmodes are localized at one
topological object. As the eigenmodes are all normalized, the maximum,
or peak, value of the eigenmode density is an indicator of how strong
this localization is. Fig.\ \ref{fig:maxHot} shows the plot of such
peak values for four eigenmodes each of three hot $16^3\times32$
configurations. This plot shows a smooth behaviour with a maximum for
$\kappa_{\mathrm{max}}\approx 0.23$ for the lowest eigenmodes. This
suggests that for hot SU(3) configurations the strongest localisation
occurs in that $\kappa$ region.

\begin{figure}[tbp]
\epsfig{angle=90, figure=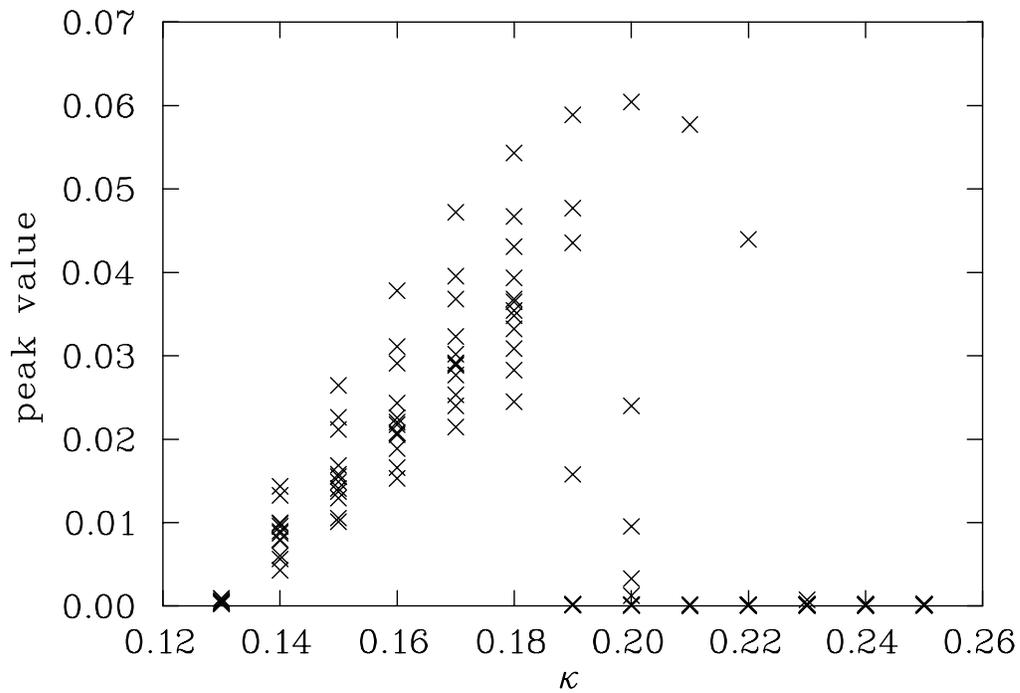, height=9.0cm}
\caption{Peak value of four eigenmode densities each from three
12-sweep cooled $16^3\times32$ configurations with respect to
$\kappa$}
\label{fig:maxCool}
\end{figure}  

Fig.\ \ref{fig:maxCool} shows the same plot for four eigenmode
densities each for three 12-sweep cooled $16^3\times32$
configurations. This plot shows a different behaviour. Again there is
a clear peak, but the values drop to almost zero immediately after the
peak. We expect this to happen after our qualitative observations in
the previous section. These low maximum values correspond to weakly
localized modes, which are on the right-hand side of the spectrum shown
in Fig.\ \ref{fig:coolSpectr}. In this case we can not determine
$\kappa_{\mathrm{max}}$, for it is not clear how the localisations are
going to develop if we follow the stronger localised eigenmodes up to
higher values of $\kappa$. The one eigenmode we could follow suggests
that $\kappa_{\mathrm{max}}\approx 0.20$. This reduction of
$\kappa_{\mathrm{max}}$ is about the same size as the reduction of
$\kappa_{\mathrm{c}}$ for going from the hot to the 12-sweep cooled
configuration.

To get more information about the shape of the eigenmodes, we fit the
instanton model, Eq.~(\ref{eq:instZeroModel}), to the eigenmodes. This
model gives us very good fits as
\begin{equation}
\sum_x (p(x)_{\mathrm{zero}}-p_{\mathrm{M}}(x))^2 = 10^{-5} \label{eq:error}
\end{equation}
where $p_{\mathrm{M}}(x)$ is the 6 parameter fit of $3^4$ points of
$p(x)$ centered about the peak of $\rho(x)$. This value is about 1000
times smaller than the peak value and $10^{-5}$ is the worst case with
most of the fits of order $10^{-7}$ to $10^{-9}$. The fit parameter
$\rho$ is then a good measurement for the size of the localization.

\begin{figure}[tbp]
\epsfig{angle=90, figure=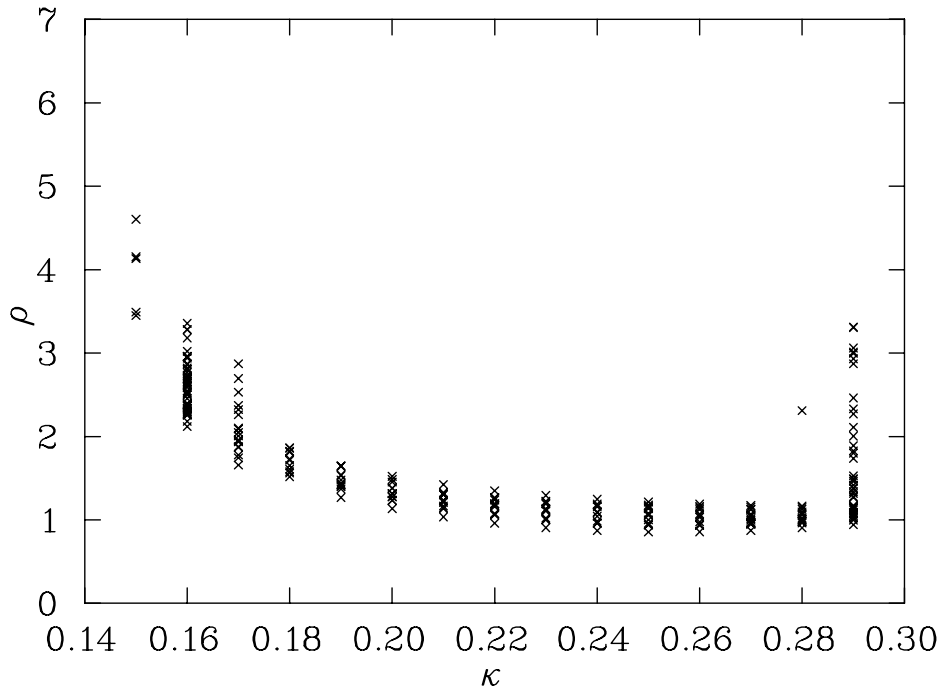, height=9.0cm}
\caption{$\rho$ of the fitted instanton model of four eigenmodes from three hot $16^3\times32$ configurations with respect to $\kappa$. Note that there are several correlations in one eigenmode at both ends of the spectrum.}
\label{fig:rhoHot}
\end{figure}

\begin{figure}[tbp]
\epsfig{angle=90, figure=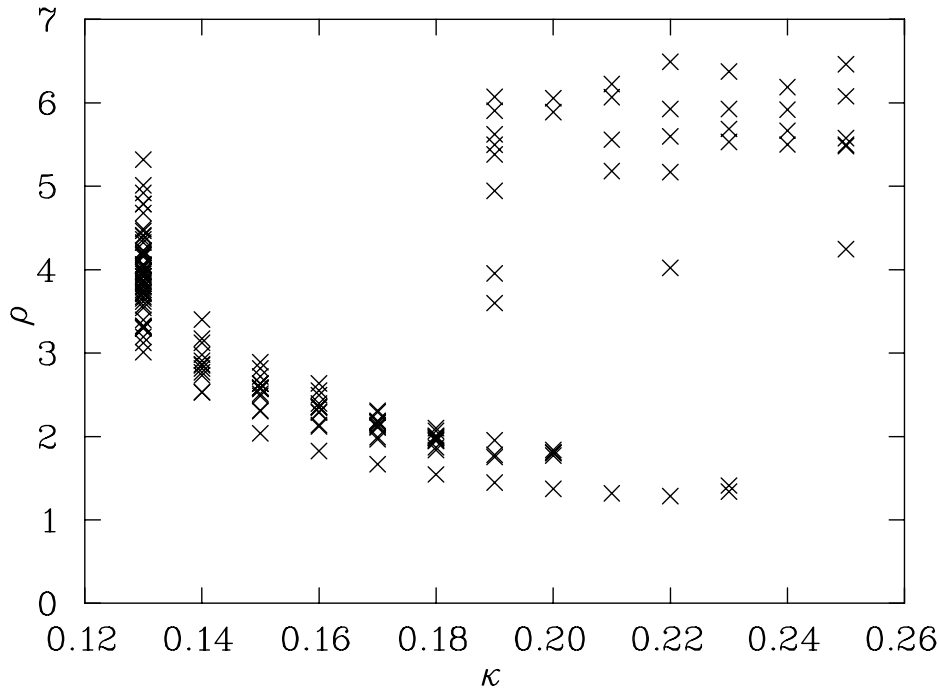, height=9.0cm}
\caption{$\rho$ of the fitted instanton model of four eigenmodes from
three 12-sweep cooled $16^3\times32$ configurations with respect to
$\kappa$. Note the jump at $\kappa \approx 0.19$.}
\label{fig:rhoCool}
\end{figure}

In Fig.\ \ref{fig:rhoHot} we plot $\rho(\kappa)$ for four eigenmodes
calculated on all hot configurations and in Fig.\ \ref{fig:rhoCool}
for four eigenmodes calculated on all 12-sweep cooled
configurations. We see a behaviour which corresponds to the behaviour
described above for the peak values of the eigenmodes. $\rho(\kappa)$
for the hot configurations shows a smooth behaviour with a minimum
around $\kappa_{\mathrm{max}} = 0.26$. This means the eigenmodes are
maximally localised for this value of $\kappa$ and are less localized
for both ends of the spectrum. The $\kappa_{\mathrm{max}}$ found this
way varies slightly from $\kappa_{\mathrm{max}}$ found by just taking
the peak values of the eigenmode densities. It is about 0.03 larger at
0.26 for the lowest eigenmode. As we mentioned earlier the eigenmodes
for low and high values of $\kappa$ are localized on more than one
topological object. Therefore Fig.\ \ref{fig:rhoHot} reports more
local maxima than modes at these $\kappa$. However in the range
$0.18\leq \kappa \leq 0.28$ only one local maximum is found per
mode.

We have established that low-lying eigenmodes are correlated to a
single topological object when eigenmodes are non degenerate. It is
also established that an instanton gives rise to a zero crossing in
the spectral flow with the sign of the slope equal to the sign of the
topological charge \cite{EHN98}. However, as mentioned above,
Fig.~\ref{fig:modes} reveals an eigenmode correlated to one
topological object, but with two zero crossings. The size of the
eigenmode varies from $\rho = 1.2$ to $\rho = 1.0$ as $\kappa$ varies
from 0.21 to 0.27. The size of the correlated object on the
topological charge density, which can only be seen on the
5-sweep-cooled configuration, is $\rho = 1.1$. We have been able to
reproduce similar spectral flows on single instanton configurations,
which have been cooled with the Wilson action to the point where the
topological charge is $Q\approx 0.4$, well below 1. Hence the double
zero crossing of the spectral flow in Fig.~\ref{fig:modes} suggests
the presence of a lattice artifact. Ideally an improved fermion action
should act to remove the zero crossings associated with this
artifact.

The behaviour of $\rho(\kappa)$ of the cooled configurations shows a
jump at $\kappa \approx 0.19$, where the weakly localised modes set in
as described above. Some of these weakly localised modes are not
sufficiently localised to allow a fit to the instanton model. A fit
would result in values for $\rho\geq 10$, which are not reasonable for
the instanton model and therefore neglected. For those eigenmodes
where we can do a fit, we get large values of $\rho$ compared to the
strongly localised eigenmodes. We find that $\rho$ is about 2 for
strongly localised eigenmodes and about 6 for weakly localised
eigenmodes. For $\kappa\leq 0.19$, $\rho(\kappa)$ decreases smoothly,
but for $\kappa\geq\ 0.19$ the values of $\rho(\kappa)$ are higher
than at the lower end of the spectrum and do not show a smooth
behaviour. Again it is hard to extract $\kappa_{\mathrm{max}}$, but it
seems that if we could follow the stronger localised modes further it
would be around 0.22. This is about 0.02 larger than the
$\kappa_{\mathrm{max}}$, which is extracted using the peak values.


The fitting of the models to the eigenmode densities, using
Eq.~(\ref{eq:instZeroModel}), as well as to the topological charge
densities, using Eq.~(\ref{eq:instActModel}), allows us to compare the
sizes for the eigenmode localisations with the sizes of the actual
topological objects. In order to do this, we find the topological
object which is located closest to the position of the eigenmode. We
can fit the instanton model only to topological charge densities of
cooled configurations, as only those are smooth enough. Therefore we
only compare the sizes of eigenmodes of cooled configurations with
sizes of actual topological objects. Due to the different localization
strength for eigenmodes on the left-hand and on the right-hand side of
the rhomboid spectrum of a cooled configuration, we look at those
eigenmodes separately.

\begin{figure}[tbp]
\epsfig{angle=90, figure=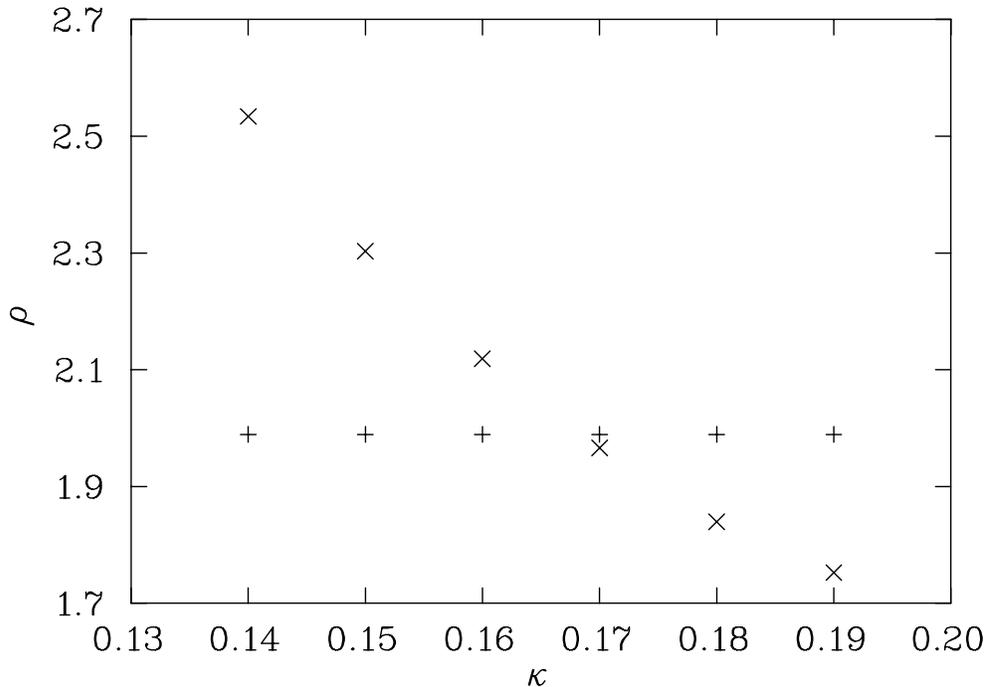, height=9.0cm}
\caption{$\rho$ of the fitted instanton model of one mode of a
12-sweep cooled $16^3\times32$ configuration with respect to
$\kappa$. The $\times$ denote the size of the eigenmode and the +
denote the size of the correlated topological object. That the two
graphs cross is a general result and all crossings are found for
$0.155 < \kappa <0.175$.}\label{fig:sizeComp}
\end{figure}

We find that the fitted positions for the strongly localised
eigenmodes agree very well. With the fitted positions of the
correlated topological objects lie within a fifth of a lattice
spacing, and most of the times even better. The fitted positions of
the weakly localised modes to the right of the spectrum agree only
within one lattice spacing with their correlated topological
objects. Strongly localised eigenmodes are correlated to smaller
objects in the topological density. Such eigenmodes on the left-hand
side of the eigenvalue spectrum, except those for
$\kappa\leq\kappa_{\mathrm{c}}$, have a size between $\rho\approx 1.5$
and $\rho\approx 3.5$. The correlated topological objects have a size
between $\rho\approx 2$ and $\rho\approx 3$. The eigenmodes are larger
than their correlated topological objects for smaller values of
$\kappa$, but as they shrink with growing $\kappa$ they get smaller
than their correlated topological objects. All followed modes reach
the size of the correlated topological object for $0.155 < \kappa
<0.175$. Fig.\ \ref{fig:sizeComp} shows an example of this
behaviour.

The weakly localised eigenmodes on the right-hand side of the
eigenvalue spectrum are larger than the strongly localised eigenmodes
on the left-hand side of the spectrum. They turn out to have a size
between $\rho\approx 4$ and $\rho\approx 6$. But the topological
charge density correlated with those eigenmodes are themselves bigger
than the topological charge density correlated with the stronger
localised modes. The sizes of the topological charge density objects
lie between $\rho\approx 3$ and $\rho \approx 4$. The weaker localised
eigenmodes are always bigger than the correlated topological objects
with $\rho_{\mathrm{mode}}\approx\rho_{\mathrm{topQ}}+2$. 

For $\kappa = \kappa_{\mathrm{c}}$ the size of the eigenmodes is
between $\rho = 3$ and $5$ with a relation to the size of the
correlated topological objects of
$\rho_{\mathrm{mode}}\approx\rho_{\mathrm{topQ}}+1$.  Sizes of
strongly localised eigenmodes of hot configurations are of a
comparable, but slightly smaller size than that on cooled
configurations.

\begin{figure}[tbp]
\epsfig{angle=90, figure=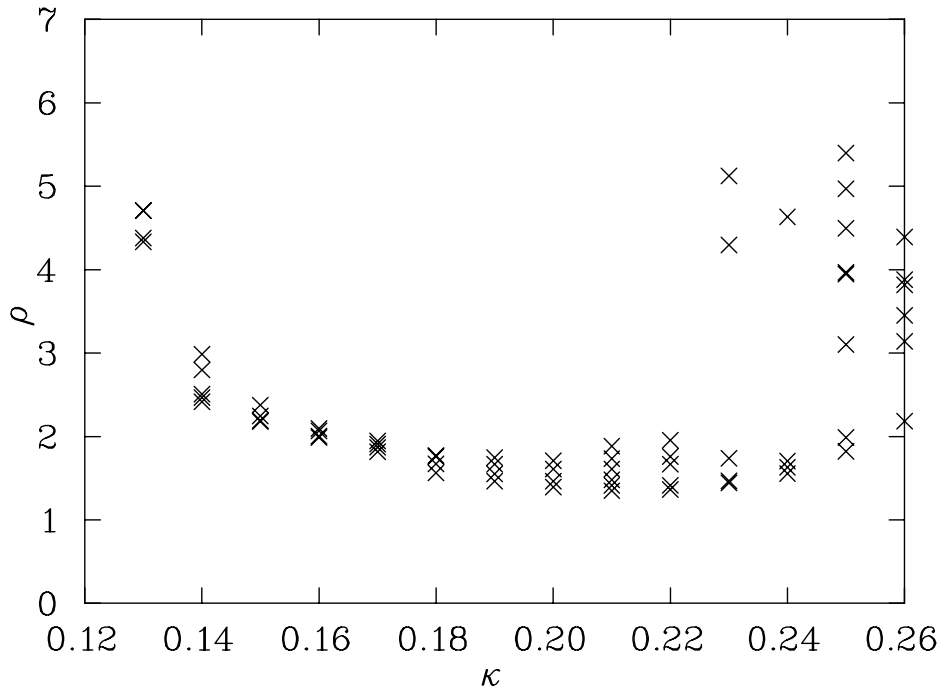, height=9.0cm}
\caption{$\rho$ of the fitted instanton model of four eigenmodes for
one 4-sweep cooled $16^3\times32$ configuration with respect to
$\kappa$. Note the jump occurring at higher $\kappa$ compared to Fig.\
\ref{fig:rhoCool}.}
\label{fig:rhoCool4}
\end{figure}

In order to understand why modes on the right-hand side of the rhomboid
are just weakly localised we calculate spectra for one configuration
with different amounts of cooling. We find that the value of $\kappa$
where the jump in the localisation size occurs becomes smaller with
cooling. Compare Fig.\ \ref{fig:rhoCool} and Fig.\ \ref{fig:rhoCool4}.

It is known \cite{EHN982} that zero crossings for larger $\kappa$
correspond to smaller topological objects. As we see localisation
sizes of calculated eigenmodes shrink with growing $\kappa$ as well,
we generally think of larger values of $\kappa$ being associated with
smaller objects. Cooling removes smaller objects first, therefore
eigenmodes for large $\kappa$ become ``unassociated'' with small
objects. The eigenmodes show a behaviour of weak localisation which
can be seen in Fig.\ \ref{fig:rhoCool} and Fig.\ \ref{fig:rhoCool4}
after the jump. Further cooling removes larger objects and therefore
the value of $\kappa$, where the change of behaviour sets in, becomes
smaller.

\section{Conclusion}\label{sect:concl}
Not only zero modes, but all low-lying eigenmodes of the hermitian
Wilson-Dirac operator $D_{\mathrm{W}}$ are strongly correlated to
topological objects for $\kappa\ge\kappa_{\mathrm{c}}$. These objects
can be instantons for which $\mathrm{S}/\mathrm{S}_0 = \mathrm{Q}$ or
topological fluctuations with $\mathrm{S}/\mathrm{S}_0 >
\mathrm{Q}$. One eigenmode is correlated to at least one topological
object with correlations to more than one object becoming more likely
as eigenvalues become degenerate and eigenmodes become broader in
size. For $\kappa<\kappa_{\mathrm{c}}$ the correlations broaden very
quickly and are lost for values smaller than about
$(\kappa_{\mathrm{c}}-0.02)$. For $\kappa>\kappa_{\mathrm{c}}$ the
correlations become sharper until $\kappa = \kappa_{\mathrm{max}}$ for
which the correlations are strongest. For
$\kappa>\kappa_{\mathrm{max}}$ the correlations broaden again.

Eigenmodes of 12-sweep cooled configurations show a different
behaviour depending on whether they belong to eigenvalues on the left
or right-hand side of the rhomboid-shaped eigenvalue spectrum of a
cooled configuration as seen in Fig.\ \ref{fig:coolSpectr}. Eigenmodes
belonging to the left-hand side of the eigenvalue spectrum are strongly
localised and show the same behaviour as eigenmodes of hot
configurations. Eigenmodes belonging to the right-hand side of the
eigenvalue spectrum are very weakly localised, but are still
correlated to topology. This suggests that those eigenmodes correspond
to very high eigenmodes in a hot configuration, which are lowered by
cooling.

The value of $\kappa$ where the weakly localised eigenmodes set in
becomes smaller with cooling. This supports the idea of high values of
$\kappa$ corresponding to localisations on small topological
objects. Small topological objects are removed first under improved
cooling thus eigenmodes for high values of $\kappa$ are the first ones
to loose the strong localisation.

When an instanton model is fitted to the eigenmode density, using
Eq.~(\ref{eq:instZeroModel}), and to the topological charge density,
using Eq.~(\ref{eq:instActModel}), strongly localised eigenmodes have
about the same size for $\rho$ as the correlated topological
objects. For $\kappa_{\mathrm{max}}$ eigenmodes are slightly smaller
than the correlated topological and for $\kappa$ smaller than
$\kappa\approx\kappa_{\mathrm{max}}-0.02 $ eigenmodes are slightly
larger.

On a single instanton configuration the correlation to the instanton
persists strongly only for the lowest eigenmode and is then gradually
lost for higher eigenmodes. On SU(3) background fields we see
correlation for the 20 lowest eigenmodes. There is only little
broadening which suggests that the correlation will persist for
eigenmodes higher than 20.

For hot configurations these correlations allow us to ``see through''
the noise to underlying topological objects. This enables one to track
the movement of these objects as a function of cooling.

Future work will examine the manner in which the eigenvalue spectrum
is modified under improved fermion actions. In particular we plan to
examine the effects of using APE-smeared fat-links in the irrelevant
operators of fermion actions.

\begin{ack}
We would like to thank the Australian National Computing Facility for
Lattice Gauge Theory for time on its supercomputer Orion and Paul
Coddington, Ramona Adorjan and Francis Vaughan for their technical
support of this work. We would like to thank Paul Rakow and Alex
Kalloniatis for useful discussions. DJK wants to thank the
Baden-W$\ddot{\mathrm{u}}$rttemberg-South Australia exchange program
for supporting his stay in Adelaide. This research was supported by
the Australian Research Council.
\end{ack}

\end{document}